\documentclass[12pt]{article}
\usepackage{amssymb}
\usepackage{epsf}
\usepackage{epsfig}
\usepackage{cite}
\newcommand{\lsim}{
\mathrel{\hbox{\rlap{\hbox{\lower4pt\hbox{$\sim$}}}\hbox{$<$}}}}

\newcommand{\gsim}{
\mathrel{\hbox{\rlap{\hbox{\lower4pt\hbox{$\sim$}}}\hbox{$>$}}}}

\newcommand{\be}{\begin{equation}}
\newcommand{\ee}{\end{equation}}
\newcommand{\bi}{\begin{itemize}}
\newcommand{\ei}{\end{itemize}}

\begin{document}
\begin{titlepage}
\vspace*{-0.5truecm}

\begin{flushright}
CERN-PH-TH/2007-044\\
TUM-HEP-663/07\\ 
UAB-FT-627\\
hep-ph/0702275
\end{flushright}

\vspace*{0.3truecm}

\begin{center}
\boldmath
{\Large{\bf 
On Puzzles and Non-Puzzles in $B\to\pi\pi,\pi K$ Decays 
}}
\unboldmath
\end{center}

\vspace{0.9truecm}

\begin{center}
{\bf Robert Fleischer,${}^a$
Stefan Recksiegel${}^b$ and Felix Schwab${}^c$}

\vspace{0.5truecm}

${}^a$ {\sl Theory Division, Department of Physics, CERN, CH-1211 Geneva 23,
Switzerland}

\vspace{0.2truecm}

${}^b$ {\sl Physik Department, Technische Universit\"at M\"unchen,
D-85748 Garching, Germany}

\vspace{0.2truecm}

  ${}^c$ {\sl Departament de F\'{\i}sica Te\`orica, IFAE, UAB,  
E-08193 Bellaterra,
  Barcelona, Spain }

\end{center}

\vspace{0.6cm}
\begin{abstract}
\vspace{0.2cm}\noindent
Recently, we have seen interesting progress in the exploration of CP violation
in $B^0_d\to\pi^+\pi^-$: the measurements of mixing-induced CP violation by the 
BaBar and Belle collaborations are now in good agreement with each other,
whereas the picture of direct CP violation is still unclear. Using the branching ratio 
and direct CP asymmetry of $B^0_d\to\pi^-K^+$, this situation can be clarified.
We predict ${\cal A}_{\rm CP}^{\rm dir}(B_d\to\pi^+\pi^-)=-0.24\pm0.04$, which 
favours the BaBar result, and extract $\gamma=\left(70.0^{+3.8}_{-4.3}\right)^\circ$, 
which agrees with the unitarity triangle fits. Extending our analysis to other 
$B\to\pi K$ modes and $B^0_s\to K^+K^-$ with the help of the $SU(3)$ flavour 
symmetry and plausible dynamical assumptions, we find that all observables with 
colour-suppressed electroweak penguin contributions are measured in excellent 
agreement with the Standard Model. As far as the ratios $R_{\rm c,n}$ of the charged 
and neutral $B\to\pi K$ branching ratios are concerned, which are sizeably affected 
by electroweak penguin contributions, our Standard-Model predictions have almost
unchanged central values, but significantly reduced errors. Since the new data have moved 
quite a bit towards these results, the ``$B\to\pi K$ puzzle" for the CP-conserving 
quantities has been significantly reduced. However, the mixing-induced CP violation of 
$B^0_d\to\pi^0K_{\rm S}$ does look puzzling; if confirmed by future measurements, 
this effect could be accommodated through a modified electroweak penguin sector 
with a large CP-violating new-physics phase. Finally, we point out that the established
difference between the direct CP asymmetries of $B^\pm\to\pi^0K^\pm$ 
and $B_d\to\pi^\mp K^\pm$ appears to be generated by hadronic and not by new 
physics.
\end{abstract}

\vspace*{0.5truecm}
\vfill
\noindent
February 2007

\end{titlepage}

\thispagestyle{empty}
\vbox{}
\newpage

\setcounter{page}{1}

\section{Introduction}\label{sec:intro}
Since many years, the $B\to\pi\pi,\pi K$ system receives a lot of attention in the
$B$-physics community (for a review, see \cite{RF-rev}). Thanks to the interplay
between theory and the $B$-factory data, valuable new insights into the physics
of these modes could be obtained, raising also the question of having an
electroweak (EW) penguin sector that is modified by the presence of CP-violating
new-physics (NP) contributions \cite{BF-00,BpiK-papers}. In this paper, we shall perform 
an analysis of the $B\to\pi\pi,\pi K$ modes in the spirit of the strategy developed in 
\cite{BFRS-I,BFRS-II}, and will, in particular, address the new data that were reported 
by the $B$-factory experiments BaBar (SLAC) and Belle (KEK) in the summer of 2006. 
The corresponding working assumptions for the treatment of the hadronic 
$B\to\pi\pi,\pi K$ sector can be summarised as follows: 
\begin{itemize}
\item[i)] {\it $SU(3)$ flavour symmetry:} $SU(3)$-breaking effects are,
however, included through ratios of decay constants and form factors 
whenever they arise, and the sensitivity of the numerical results on 
non-factorizable $SU(3)$-breaking effects is explored.
\item[ii)] {\it Neglect of the penguin annihilation and exchange topologies:} these 
contributions can be probed and controlled through the $B_d\to K^+K^-$, 
$B_s\to\pi^+\pi^-$ system \cite{GHLR,BFRS-I}, which can be fully exploited 
at the LHCb (CERN) experiment.
\end{itemize}
The data support these hypotheses, as all consistency checks that can 
currently be performed do not indicate any anomalous behaviour. The following
analysis is essentially a study within the Standard Model (SM), with the goal
to perform tests of the Kobayashi--Maskawa (KM) mechanism of CP violation 
\cite{KM}. However, also the effects of NP can straightforwardly be explored if we
assume that it manifests itself only in the EW penguin sector. This scenario was,
on the one hand, driven by the comparison of the $B$-factory data for the $B\to\pi K$
observables with their SM predictions. On the other hand, such a kind of physics 
beyond the SM can also be accommodated in various specific frameworks, 
including supersymmetry, models with  extra $Z'$ bosons, and scenarios with 
extra dimensions.

The outline of this paper is as follows: the starting point of our analysis, the
$B\to\pi\pi$ system, is discussed in Section~\ref{sec:Bpipi}. In Section~\ref{sec:BpiK},
we then move on to the $B\to\pi K$ decays. Finally, our main conclusions and a
brief outlook are given in Section~\ref{sec:concl}.

\boldmath
\section{The $B\to\pi\pi$ System}\label{sec:Bpipi}
\unboldmath
The $B\to\pi\pi$ system consists of the decays $B^0_d\to\pi^+\pi^-$, $B^+\to\pi^+\pi^0$
and $B^0_d\to\pi^0\pi^0$ and their charge conjugates. As is well known, the 
corresponding decay amplitudes can be related to one another with the help of the 
$SU(2)$ isospin symmetry \cite{GL}, which allows us also to take the effects of EW 
penguin topologies into account \cite{BF98,GPY}. We shall come back to this feature 
below. 

\boldmath
\subsection{CP Violation in $B^0_d\to\pi^+\pi^-$}
\unboldmath
Let us first have a closer look at the CP violation in the $B^0_d\to\pi^+\pi^-$ channel, 
which receives contributions from tree and penguin diagrams in the SM. The 
corresponding decay amplitude can be written in the following form \cite{RF-Bpipi}:
\begin{equation}\label{Api+pi-}
A(B^0_d\to\pi^+\pi^-)=-|\tilde T| e^{i\delta_{\tilde T}}
\left[e^{i\gamma}-de^{i\theta}\right],
\end{equation}
where the $\tilde T$ amplitude is governed by the colour-allowed tree topologies, 
$\gamma$ is the usual angle of the unitarity triangle, and the CP-conserving 
hadronic parameter $de^{i\theta}$ describes -- sloppily speaking -- the ratio of penguin
to tree contributions. The interference between the different weak amplitudes 
in (\ref{Api+pi-}) leads to a {\it direct} CP asymmetry
${\cal A}^{\rm dir}_{\rm CP}(B_d\to\pi^+\pi^-)$, whereas the interference between 
$B^0_d$--$\bar B^0_d$ mixing and the $B^0_d,\bar B^0_d\to\pi^+\pi^-$ decay 
processes generates a {\it mixing-induced} CP asymmetry 
${\cal A}^{\rm mix}_{\rm CP}(B_d\to\pi^+\pi^-)$. These observables enter
the following time-dependent CP asymmetry \cite{RF-rev}:
\begin{eqnarray}
\lefteqn{\frac{\Gamma(B^0_d(t)\to \pi^+\pi^-)-
\Gamma(\bar B^0_d(t)\to \pi^+\pi^-)}{\Gamma(B^0_d(t)\to \pi^+\pi^-)+
\Gamma(\bar B^0_d(t)\to \pi^+\pi^-)}}\nonumber\\
&&={\cal A}_{\rm CP}^{\rm dir}(B_d\to\pi^+\pi^-)\,\cos(\Delta M_d t)+
{\cal A}_{\rm CP}^{\rm mix}(B_d\to\pi^+\pi^-)\,\sin(\Delta M_d t).\label{ACP-timedep}
\end{eqnarray}
As in \cite{BFRS-I,BFRS-II}, we shall use a sign convention similar to that of 
(\ref{ACP-timedep}) also for self-tagging neutral $B_d$ and charged $B$ decays\footnote{Our definition of ${\cal A}_{\rm CP}^ {\rm dir}$ coincides with
$C_{CP}=-A_{CP}$ but our ${\cal A}_{\rm CP}^{\rm mix}$ differs in sign from $S_{CP}$ \cite{HFAG}}.

Concerning the measurement of CP violation in $B^0_d\to\pi^+\pi^-$, there has
been interesting recent progress. There is now -- for the first time -- a nice 
agreement between the BaBar and Belle results for the mixing-induced 
CP asymmetry:
\begin{equation}
{\cal A}_{\rm CP}^{\rm mix}(B_d\to\pi^+\pi^-)=
\left\{\begin{array}{ll}
0.53\pm0.14\pm0.02 & \mbox{(BaBar \cite{BaBar-Bpi+pi-})}\\
0.61\pm0.10\pm0.04 & \mbox{(Belle \cite{Belle-Bpi+pi-}),}
\end{array}\right.
\end{equation}
which yields the average of ${\cal A}_{\rm CP}^{\rm mix}(B_d\to\pi^+\pi^-)=
0.59\pm0.09$ \cite{HFAG}.
On the other hand, the picture of direct CP violation is still {\it not} settled:
\begin{equation}\label{ACP-dir-pipi-ex}
{\cal A}_{\rm CP}^{\rm dir}(B_d\to\pi^+\pi^-)=\left\{
\begin{array}{cc}
-0.16\pm0.11\pm0.03 & \mbox{(BaBar \cite{BaBar-Bpi+pi-})}\\
-0.55\pm0.08\pm0.05 & \mbox{(Belle \cite{Belle-Bpi+pi-}).}
\end{array}\right.
\end{equation}

\begin{figure} 
   \begin{center}
   \includegraphics[width=8.5truecm]{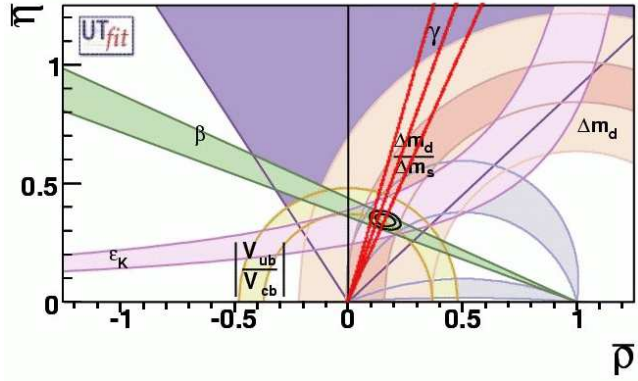} 
  \end{center}
  \vspace*{-0.5truecm}
   \caption{Comparison of the value of $\gamma$ in (\ref{gam}) with the
   SM fit of the unitarity triangle obtained by the UTfit collaboration \cite{UTfit}; 
   the situation is analogous for the corresponding analysis of the CKMfitter 
   collaboration \cite{CKMfitter}.}\label{fig:UTfit}
\end{figure}

\boldmath
\subsection{Clarifying the Picture through $B^0_d\to\pi^-K^+$}\label{ssec:clarify}
\unboldmath
This unsatisfactory situation can be resolved with the help of the $B^0_d\to\pi^-K^+$
mode, which receives --  in analogy to the $B^0_d\to\pi^+\pi^-$ channel -- also contributions from tree and penguin topologies. However, since $B^0_d\to\pi^-K^+$
is caused by $\bar b\to \bar s u\bar u$ quark-level transitions, it exhibits an 
amplitude hierarchy which is different from that of the $\bar b\to \bar d u\bar u$ decay
$B^0_d\to\pi^+\pi^-$, and is actually dominated by the QCD penguin topologies.
Direct CP violation in this decay, which is generated through the interference
between the penguin and tree contributions, is now experimentally well 
established:
\begin{equation}
{\cal A}_{\rm CP}^{\rm dir}(B_d\to\pi^\mp K^\pm)=
\left\{\begin{array}{ll}
 0.108\pm0.024\pm0.008 &  \mbox{(BaBar \cite{BaBar-Bpi+pi-})}\\
 0.093\pm0.018\pm0.008 &  \mbox{(Belle \cite{Belle-ACP-Bpi-K+})}\\
0.04\pm0.16\pm0.02& \mbox{(CLEO \cite{CLEO-BpiK})}\\
0.086\pm0.023\pm0.009 & \mbox{(CDF \cite{CDF-punzi}),}
\end{array}\right.
\end{equation}
leading to the average of 
${\cal A}_{\rm CP}^{\rm dir}(B_d\to\pi^\mp K^\pm)=0.095\pm0.013$ \cite{CDF-punzi}.
In the SM, we may write
\begin{equation}
A(B^0_d\to \pi^- K^+)=P'\left[1-re^{i\delta}e^{i\gamma}\right],
\end{equation}
where the penguin amplitude $P'$ and $re^{i\delta}$, which measures -- sloppily 
speaking -- the ratio of tree to penguin contributions, are CP-conserving strong
quantities. Using the $SU(3)$ flavour symmetry and the dynamical assumptions 
specified in Section~\ref{sec:intro}, we obtain 
\begin{equation}
re^{i\delta}=\frac{\epsilon}{d}e^{i(\pi-\theta)},
\end{equation}
where $\epsilon\equiv\lambda^2/(1-\lambda^2)=0.05$ involves the usual Wolfenstein
parameter \cite{wolf}, which implies the following
relation \cite{RF-Bpipi,DHGR}:
\begin{equation}
H_{\rm BR}\equiv
\frac{1}{\epsilon}\left(\frac{f_K}{f_\pi}\right)^2\left[\frac{\mbox{BR}
(B_d\to\pi^+\pi^-)}{\mbox{BR}(B_d\to\pi^\mp K^\pm)}\right]
=-\frac{1}{\epsilon}\left[\frac{{\cal A}_{\rm CP}^{\rm dir}(B_d\to\pi^\mp 
K^\pm)}{{\cal A}_{\rm CP}^{\rm dir}(B_d\to\pi^+\pi^-)}\right].
\end{equation}
Since the CP-averaged branching ratios and the direct CP 
violation in $B^0_d\to\pi^-K^+$ are well measured \cite{HFAG}, we may use this 
relation to {\it predict} the direct CP asymmetry
\begin{equation}
{\cal A}_{\rm CP}^{\rm dir}(B_d\to\pi^+\pi^-)=-0.24\pm0.04,
\end{equation}
which favours the BaBar result in (\ref{ACP-dir-pipi-ex}). Furthermore, as
the $B^0_d$--$\bar B^0_d$ mixing phase 
\begin{equation}\label{phid}
\phi_d=(42.4\pm2)^\circ
\end{equation}
is known through the CP violation measurements in $B\to J/\psi K^{(*)}$ \cite{HFAG}, 
the quantities $H_{\rm BR}$, ${\cal A}_{\rm CP}^{\rm dir}(B_d\to\pi^\mp K^\pm)$ and 
${\cal A}_{\rm CP}^{\rm mix}(B_d\to\pi^+\pi^-)$ can be expressed in terms of 
$\gamma$ and $d$, $\theta$. Consequently, these parameters can be extracted 
from the data. As far as the angle $\gamma$ is concerned, we obtain
\begin{equation}\label{gam}
\gamma=\left(70.0^{+3.8}_{-4.3}\right)^\circ,
\end{equation}
which is in nice agreement with the SM fits of the unitarity triangle \cite{UTfit,CKMfitter},
as can be seen in Fig.~\ref{fig:UTfit}. For the remainder of this analysis, we will use
the value of $\gamma$ in (\ref{gam}).

\boldmath
\subsection{Hadronic Parameters and CP Violation in $B^0_d\to\pi^0\pi^0$}
\unboldmath
Concerning the determination of the ratio of the penguin to tree amplitudes of 
the $B^0_d\to\pi^+\pi^-$ decay as described above, we find
\begin{equation}\label{d-theta}
d=0.46\pm0.02,\quad \theta=(155\pm4)^\circ.
\end{equation}
The $B\to\pi\pi$ system offers two more channels. Using the isospin symmetry of
strong interactions, their decay amplitudes can be written as follows \cite{BFRS-I}:
\begin{eqnarray}
\sqrt{2}A(B^+\to\pi^+\pi^0)&=&-|\tilde T|e^{i\delta_{\tilde T}}e^{i\gamma}
\left[1+xe^{i\Delta}\right]\\
\sqrt{2}A(B^0_d\to\pi^0\pi^0)&=&|P|e^{i\delta_P}
\left[1+(x/d)e^{i\gamma}e^{i(\Delta-\theta)}\right],\label{Api0pi0}
\end{eqnarray}
where the hadronic parameter $xe^{i\Delta}$ denotes the ratio of 
``colour-suppressed" to ``colour-allowed tree'' amplitudes. Since we have
two more $B\to\pi\pi$ observables at our disposal, 
\begin{eqnarray}
R_{+-}^{\pi\pi}&\equiv&
2\left[\frac{\mbox{BR}(B^\pm\to\pi^\pm\pi^0)}{\mbox{BR}
(B_d\to\pi^+\pi^-)}\right]\frac{\tau_{B^0_d}}{\tau_{B^+}} = 2.02\pm0.16\\
R_{00}^{\pi\pi}&\equiv&2\left[\frac{\mbox{BR}(B_d\to\pi^0\pi^0)}{\mbox{BR}
(B_d\to\pi^+\pi^-)}\right] = 0.50\pm0.08,
\end{eqnarray}
where we have also given the most recent experimental averages \cite{HFAG}, 
$x$ and the strong phase $\Delta$ can be determined:
\begin{equation}\label{x-Delta}
x=0.92_{-0.09}^{+0.08},\quad \Delta=-(50_{-14}^{+11})^\circ.
\end{equation}

If we use (\ref{Api0pi0}) and complement (\ref{x-Delta}) with the numbers in (\ref{phid}), 
(\ref{gam}) and (\ref{d-theta}), the CP asymmetries of the $B^0_d\to\pi^0\pi^0$ 
channel can be {\it predicted} in the SM. Following these lines, we obtain the numbers 
\begin{eqnarray}
{\cal A}_{\rm CP}^{\rm dir}(B_d\to\pi^0\pi^0)|_{\rm SM}
&=&-(0.40^{+0.14}_{-0.21})\label{ACPdir00}\\
{\cal A}_{\rm CP}^{\rm mix}(B_d\to\pi^0\pi^0)|_{\rm SM}&=&
-(0.71^{+0.16}_{-0.17}),
\end{eqnarray}
which offer the exciting perspective of observing {\it large} CP violation in this
decay. So far, only data for the direct CP asymmetry are available from the 
BaBar \cite{BaBar-Bpi+pi-} and Belle \cite{Belle-ACPpi0pi0} collaborations, 
yielding the following average \cite{HFAG}:
\begin{equation}
{\cal A}_{\rm CP}^{\rm dir}(B_d\to\pi^0\pi^0)=-(0.36^{+0.33}_{-0.31}).
\end{equation}
Although this result is still compatible with zero at the 1.1 $\sigma$ level, 
the agreement with  (\ref{ACPdir00}) (note the sign) is nevertheless very
encouraging and gives us further confidence in our analysis. 
Let us finally note that also 
the EW penguin contributions are included in our numerical values
with the help of the isospin symmetry \cite{BF98,GPY,BFRS-II}, 
although they have a tiny impact on the $B\to\pi\pi$ system.

\boldmath
\section{The $B\to\pi K$ System}\label{sec:BpiK}
\unboldmath
Let us now turn to our main target, which is given by the $B\to\pi K$ system. In 
addition to the $B^0_d\to\pi^-K^+$ mode, it consists of $B^+\to\pi^+K^0$, 
$B^0_d\to\pi^0K^0$ and $B^+\to\pi^0K^+$, as well as their charge conjugates. 
In the SM, all these decays are governed by their QCD penguin contributions. 
However, also EW penguins may play an important r\^ole. We distinguish between 
the following cases:
\begin{itemize}
\item {\it Colour-suppressed} EW penguins with tiny effects:
$B^0_d\to\pi^-K^+$, $B^+\to\pi^+K^0$.
\item {\it Colour-allowed} EW penguins with sizeable effects:
$B^0_d\to\pi^0K^0$, $B^+\to\pi^0K^+$.
\end{itemize}
In the latter case, the EW penguin contributions are even comparable to those
of colour-allowed tree topologies. Let us first have a closer look at those $B\to\pi K$ observables that are marginally affected by EW penguins.

\subsection{Tiny Electroweak Penguin Effects}
\boldmath
\subsubsection{$B^0_d\to\pi^-K^+$ and $B^+\to\pi^+K^0$}\label{sssec:BpiK-mixed}
\unboldmath
In Subsection~\ref{ssec:clarify}, we have already used the CP-averaged branching 
ratio and the direct CP asymmetry of the $B^0_d\to\pi^- K^+$ channel. As we saw
in Fig.~\ref{fig:UTfit}, the resulting value of $\gamma$ in (\ref{gam}) agrees nicely 
with the SM fits of the unitarity triangle. However, there is another decay with
tiny (colour-suppressed) EW penguin contributions at our disposal, the $B^+\to\pi^+K^0$
channel. In the SM, its decay amplitude can be written as
\begin{equation}
A(B^+\to\pi^+K^0)=-P'\left[1+\rho_{\rm c}e^{i\theta_{\rm c}}
e^{i\gamma}\right],
\end{equation}
where the CP-conserving hadronic parameter $\rho_{\rm c}e^{i\theta_{\rm c}}$ is
doubly Cabibbo-suppressed and, hence, usually neglected. In this limit, we obviously 
have vanishing direct CP violation in $B^+\to\pi^+K^0$. This feature is fully supported 
by the following experimental average \cite{HFAG}:
\begin{equation}
{\cal A}_{\rm CP}^{\rm dir}(B^\pm\to\pi^\pm K)=-0.009\pm0.025.
\end{equation}

Finally, using the working assumptions specified in Section~\ref{sec:intro}, we can
predict the following ratio of CP-averaged branching ratios \cite{FM}:
\begin{eqnarray}
R  & \equiv & \left[\frac{\mbox{BR}(B^0_d\to\pi^- K^+)+
\mbox{BR}(\bar B^0_d\to\pi^+ K^-)}{\mbox{BR}(B^+\to\pi^+ K^0)+
\mbox{BR}(B^-\to\pi^- \bar K^0)}\right]\frac{\tau_{B^+}}{\tau_{B^0_d}}\nonumber\\
& & \stackrel{\rm SM}{=}  0.942\pm0.012 \stackrel{\rm exp}{=}0.93\pm0.05.
\end{eqnarray}
The excellent agreement of our SM prediction with the data is impressive, and in 
particular no anomalous value of $\rho_{\rm c}$ is indicated. In this context it is 
interesting to note that a similar picture of $\rho_{\rm c}$ follows also from the recently 
observed $B^\pm\to K^\pm K$ decays \cite{FR}. Consequently, toy models of 
final-state interaction effects yielding a significant enhancement of the 
$\rho_{\rm c}$ parameter that were discussed several years ago are now strongly disfavoured
out by the $B$-factory data. 

\boldmath
\subsubsection{Another Application: Prediction of  the $B^0_s\to K^+K^-$ Observables}
\unboldmath
As an interesting by-product, the strategy developed in \cite{BFRS-I,BFRS-II} allows us 
to predict the observables of the $B^0_s\to K^+K^-$ decay, where the 
(colour-suppressed) EW penguin contributions have again a tiny impact. Using the SM 
value of the $B^0_s$--$\bar B^0_s$ mixing phase $\phi_s=-2\lambda^2\eta=-2^\circ$ 
\cite{RF-rev}, where $\lambda$ and $\eta$ are the usual parameters of the 
Wolfenstein parametrization \cite{wolf}, we arrive at the following predictions of the 
CP asymmetries:
\begin{eqnarray}
{\cal A}_{\rm CP}^{\rm dir}(B_s\to K^+K^-)|_{\rm SM}&=&
0.093\pm0.015\\
{\cal A}_{\rm CP}^{\rm mix}(B_s\to K^+K^-)|_{\rm SM}&=&-0.234_{-0.014}^{+0.017}.
\end{eqnarray}
In the case of the CP-averaged branching ratio, an $SU(3)$-breaking form-factor
ratio enters the prediction, thereby increasing the uncertainties. If we use the
result of a QCD sum-rule calculation \cite{Khod}, we obtain
\begin{equation}\label{BsKK-BR}
\mbox{BR}(B_s\to K^+K^-)=(28_{-5}^{+7})\times 10^{-6}.
\end{equation}
The $B^0_s\to K^+K^-$ channel was recently observed by the CDF
collaboration \cite{CDF-BsK+K-}; the most recent experimental result for the
CP-averaged branching ratio reads as follows \cite{CDF-punzi}: 
\begin{equation}\label{BsKK-exp}
\mbox{BR}(B_s\to K^+K^-)=(24.4\pm1.4\pm4.6)\times10^{-6}.
\end{equation}
Within the uncertainties, (\ref{BsKK-BR}) is in nice agreement with (\ref{BsKK-exp}),
which is another support of the working hypotheses listed in Section~\ref{sec:intro}.
The $B_s\to K^+K^-$, $B_d\to\pi^+\pi^-$ system offers a strategy for the extraction 
of $\gamma$ with the help of the $U$-spin flavour symmetry of strong interactions
\cite{RF-Bpipi}, which can nicely be implemented at the LHCb experiment \cite{nardulli}.
The predictions and hadronic parameters given above are useful for further 
experimental studies in the preparation for the quickly approaching start of the LHC.

\subsection{Sizeable Electroweak Penguin Effects}
\subsubsection{CP-Conserving Observables}
Let us now focus on those CP-conserving $B\to\pi K$ observables that are sizeably 
affected by EW penguin contributions. In this context, the following ratios 
\cite{BF98} have received a lot of attention in the literature:
\begin{eqnarray}
R_{\rm c}&\equiv&2\left[\frac{\mbox{BR}(B^+\to\pi^0K^+)+
\mbox{BR}(B^-\to\pi^0K^-)}{\mbox{BR}(B^+\to\pi^+ K^0)+
\mbox{BR}(B^-\to\pi^- \bar K^0)}\right]=1.11\pm0.07\\
R_{\rm n}&\equiv&\frac{1}{2}\left[
\frac{\mbox{BR}(B_d^0\to\pi^- K^+)+\mbox{BR}(\bar B_d^0\to\pi^+ 
K^-)}{\mbox{BR}(B_d^0\to\pi^0K^0)+\mbox{BR}(\bar B_d^0\to\pi^0\bar K^0)}
\right]=0.99\pm0.07,
\end{eqnarray}
where we have also given the experimental averages \cite{HFAG}, taking
the most recent measurements by the BaBar \cite{BaBar-Bpi+pi-} and Belle 
\cite{Belle-BpiK} collaborations into account.
In these quantities, the EW penguin effects enter in colour-allowed form through 
the modes involving neutral pions, and are theoretically described by a parameter 
$q$, which measures the ``strength" of the EW penguin with respect to the tree 
contributions, and a CP-violating phase $\phi$. In the SM, the $SU(3)$ flavour 
symmetry allows a prediction of $q=0.60$ \cite{NR}, and $\phi$ 
{\it vanishes.} As is known for many years (see, for instance, \cite{EWP-NP}), 
EW penguin topologies offer an interesting avenue for NP to manifest itself in 
the $B$-factory data. In the case of CP-violating NP effects of this kind, $\phi$
would take a value different from zero.

\begin{figure}
\begin{center}
\includegraphics[width=0.6\textwidth]{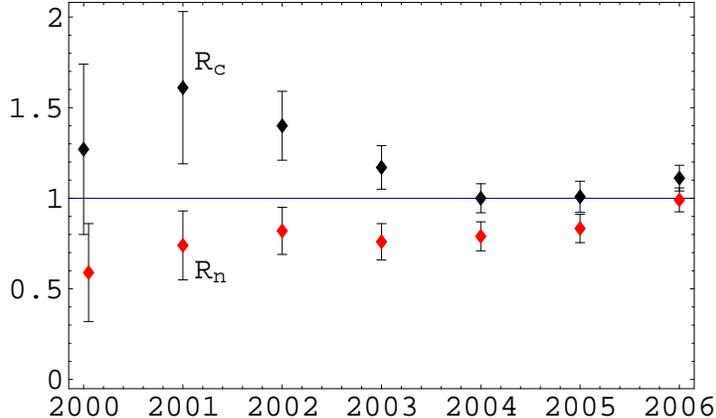}
\end{center}
\vspace*{-0.5truecm}
\caption{\label{fig:timeline}The time evolution of the experimental values of 
$R_{\rm c}$ and $R_{\rm n}$.}
\end{figure}

\begin{figure}
\begin{center}
\includegraphics[width=0.7\textwidth]{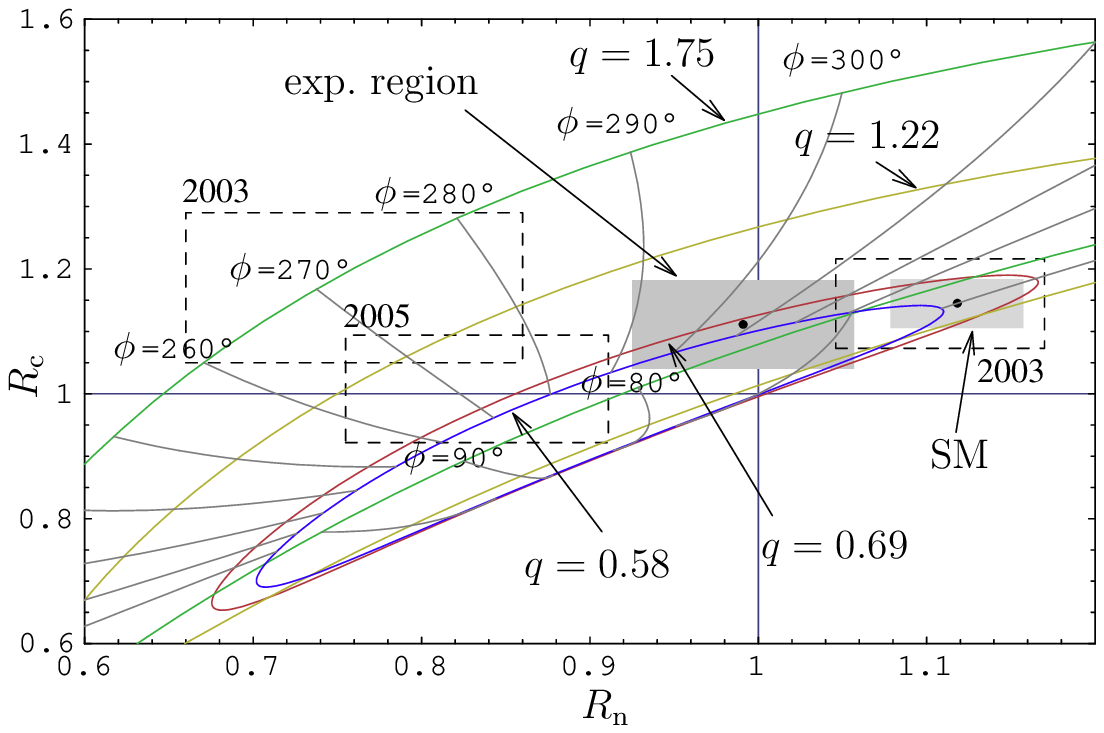}
\end{center}
\vspace*{-0.5truecm}
\caption{\label{fig:RnRc} The situation in the $R_{\rm n}$--$R_{\rm c}$ plane,
as discussed in the text.}
\end{figure}

As can be seen in Fig.~\ref{fig:timeline}, which illustrates the time evolution of
the measurements of $R_{\rm c}$ and $R_{\rm n}$, the central values have 
significantly moved up with respect to the 2005 data
(partly due to radiative corrections affecting final states 
with charged particles \cite{BarIsi}), while the errors were only marginally reduced.
Following \cite{BFRS-I,BFRS-II}, let us now discuss the situation in the plane of 
$R_{\rm n}$ and $R_{\rm c}$, as shown in Fig.~\ref{fig:RnRc}. Here the various
contours correspond to different values of $q$, and the position on the contour is 
parametrized through the CP-violating phase $\phi$. We observe that the SM
prediction (on the right-hand side) is very stable in time, having now significantly
reduced errors. On the other hand, the $B$-factory data have moved quite
a bit towards the SM, thereby reducing the ``$B\to\pi K$ puzzle" for the CP-averaged
branching ratios. A similar trend is also seen in the measurements of CP violation 
in $b\to s$ penguin-dominated decays \cite{HFAG}, where in particular the average 
value of $(\sin2\beta)_{\phi K_{\rm S}}$ has moved towards the reference value of
$(\sin2\beta)_{J/\psi K_{\rm S}}$. If we convert the experimental values of
$R_{\rm n}$ and $R_{\rm c}$ into $q$ and $\phi$, we obtain
\begin{equation}
q = 0.65_{-0.35}^{+0.39},\quad
\phi = -(52^{+21}_{-50})^\circ. 
\end{equation}
In comparison with the situation of the ratio $R$ discussed 
in Subsection~\ref{sssec:BpiK-mixed}, the agreement between the new data for
the $R_{\rm c,n}$ and their SM predictions is not as perfect. However, a case
for a modified EW penguin sector cannot be made through the new measurements
of these quantities.

\subsubsection{CP-Violating Observables}
\begin{figure}
\begin{center}
\includegraphics[width=0.7\textwidth]{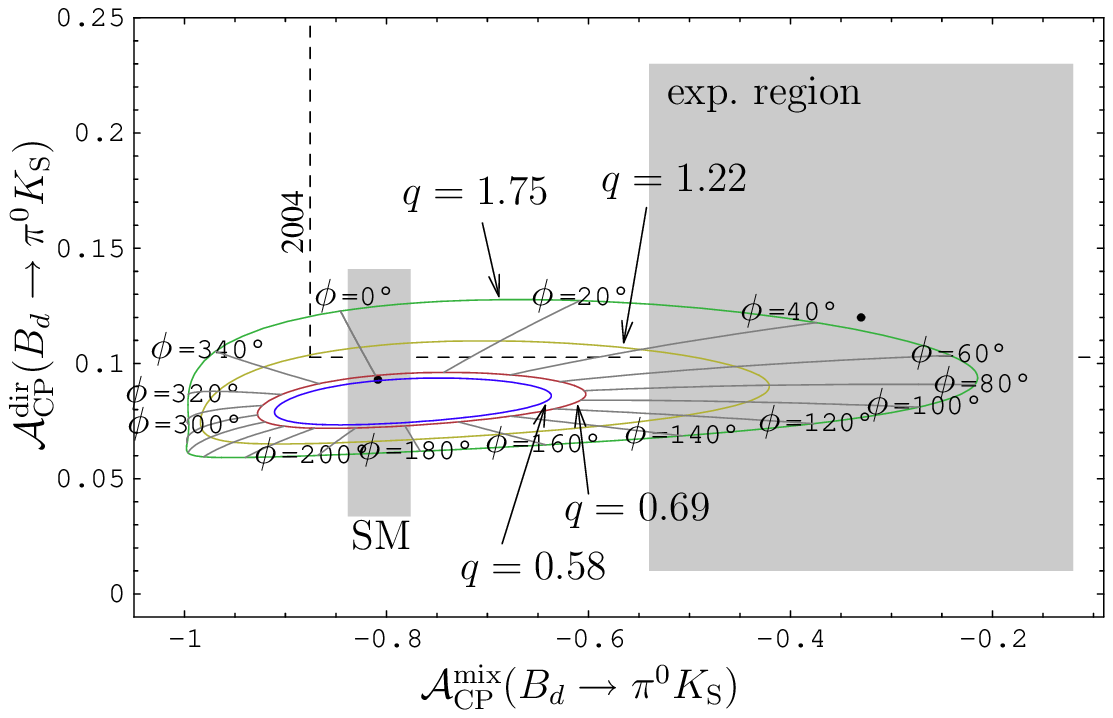}
\end{center}
\vspace*{-0.5truecm}
\caption{\label{fig:ACP}The situation in the 
${\cal A}_{\rm CP}^{\rm mix}(B_d\to\pi^0K_{\rm S})$--${\cal A}_{\rm CP}^{\rm 
dir}(B_d\to\pi^0K_{\rm S})$ plane.}
\end{figure}

In addition to the CP-conserving observables discussed above, we can also
analyse the CP-violating asymmetries of the $B^0_d\to\pi^0 K_{\rm S}$ and 
$B^\pm\to\pi^0K^\pm$ channels \cite{BFRS-I,BFRS-II}. Let us first turn to the neutral
decay, which offers an interesting probe for NP \cite{PAP-III}. Within the SM, 
we obtain the following predictions:
\begin{eqnarray}
{\cal A}_{\rm CP}^{\rm dir}(B_d\to\pi^0K_{\rm S})|_{\rm SM}&=&0.091^{+0.048}_{-0.059}\\
{\cal A}_{\rm CP}^{\rm mix}(B_d\to\pi^0K_{\rm S})|_{\rm SM}&=&-0.81\pm0.03,
\end{eqnarray}
which are much sharper than the current $B$-factory data:
\begin{equation}
{\cal A}_{\rm CP}^{\rm dir}(B_d\to\pi^0K_{\rm S})=
\left\{\begin{array}{ll}
0.20 \pm 0.16 \pm 0.03 & \mbox{(BaBar \cite{BaBar-Bpi0KS})}\\
0.05 \pm 0.14 \pm 0.05  & \mbox{(Belle \cite{Belle-Bpi0KS})}
\end{array}\right.
\end{equation}
\begin{equation}
{\cal A}_{\rm CP}^{\rm mix}(B_d\to\pi^0K_{\rm S})=
\left\{\begin{array}{ll}
-0.33 \pm 0.26 \pm 0.04 & \mbox{(BaBar \cite{BaBar-Bpi0KS})}\\
-0.33 \pm 0.35 \pm 0.08 & \mbox{(Belle \cite{Belle-Bpi0KS}),}
\end{array}\right.
\end{equation}
yielding the following averages \cite{HFAG}: 
\begin{equation}
{\cal A}_{\rm CP}^{\rm dir}(B_d\to\pi^0K_{\rm S})=0.12 \pm 0.11, \quad
{\cal A}_{\rm CP}^{\rm mix}(B_d\to\pi^0K_{\rm S})=-0.33 \pm 0.21.
\end{equation}
In analogy to Fig.~\ref{fig:RnRc}, we show the situation in the 
${\cal A}_{\rm CP}^{\rm mix}(B_d\to\pi^0K_{\rm S})$--${\cal A}_{\rm CP}^{\rm dir}
(B_d\to\pi^0K_{\rm S})$ plane in Fig.~\ref{fig:ACP}. We see
that ${\cal A}_{\rm CP}^{\rm mix}(B_d\to\pi^0K_{\rm S})$ offers a particularly 
interesting observable, and that the experimental central values can be reached 
for large {\it positive} values of $\phi$. 

Concerning direct CP violation in $B^\pm\to\pi^0K^\pm$, we obtain the following
prediction:
\begin{equation}
{\cal A}^{\rm dir}_{\rm CP}(B^\pm\to\pi^0K^\pm)|_{\rm SM}=
-0.001^{+0.049}_{-0.041},
\end{equation}
which is in good agreement with the experimental average \cite{HFAG}
\begin{equation}
{\cal A}^{\rm dir}_{\rm CP}(B^\pm\to\pi^0K^\pm)\stackrel{\rm exp}{=}-0.047\pm0.026
\end{equation}
within the errors. For the new input data, this feature turns out to 
be almost independent of the presence of CP-violating NP contributions to the
EW penguin sector. Consequently, the non-vanishing experimental
value of 
\begin{equation}\label{Delta-A}
\Delta A \equiv {\cal A}_{\rm CP}^{\rm dir}(B^\pm\to\pi^0K^\pm)-
{\cal A}_{\rm CP}^{\rm dir}(B_d\to\pi^\mp K^\pm)
\stackrel{\rm exp}{=}-0.140\pm0.030,
\end{equation}
which differs from zero at the $4.7\,\sigma$ level, is likely to be generated through hadronic effects, i.e.\ not through the impact of physics beyond the SM. A similar
conclusion was drawn in \cite{GR-06}, where it was also noted that the measured
values of $R_{\rm c}$ and $R_{\rm n}$ are now in accordance with the SM.

Finally, performing a simultaneous fit to $R_{\rm n}$, $R_{\rm c}$ and the 
CP-violating observables of $B^0_d\to\pi^0K_{\rm S}$, we arrive at
\begin{equation}
q=1.7_{-1.3}^{+0.5},\quad \phi=+\left(73_{-18}^{+6}\right)^\circ.
\end{equation}
Interestingly, these parameters -- in particular the large {\it positive} phase -- 
would also allow us to accommodate the experimental values of
$(\sin2\beta)_{\phi K_{\rm S}}$ and the CP asymmetries of other 
$b\to s$ penguin modes with central values smaller than 
$(\sin2\beta)_{\psi K_{\rm S}}$. The large value of $q$ would be excluded 
by constraints from rare decays in simple scenarios where NP enters only 
through $Z$ penguins \cite{BFRS-I,BFRS-II}, but could still be accommodated in other 
scenarios, e.g.\ in models with leptophobic $Z'$ bosons.

\boldmath
\subsection{Sensitivity on $SU(3)$-Breaking Effects}
\unboldmath
In analogy to the detailed discussion in \cite{BFRS-II}, we have also explored the
sensitivity of our new numerical results on $SU(3)$-breaking effects. 
The resulting picture is essentially the same: Even if we allow
for very conservative uncertainties in the $SU(3)$-breaking effects, the
resulting uncertainties on our predictions are only very moderately
increased with respect to our standard analysis, the $SU(3)$-breaking
corrections never become dominant over the other sources of uncertainty.

For numerical details, we refer the reader to the discussion in \cite{BFRS-II}.
We would like to stress that even our standard analysis takes into
account a reasonable estimate of $SU(3)$-breaking related uncertainties
which are included in all the theoretical errors quoted in this paper.

\section{Conclusions and Outlook}\label{sec:concl}
The $B\to\pi\pi,\pi K$ system remains a particularly interesting playground for
the testing of the KM mechanism of CP violation, and the systematic strategy 
developed in \cite{BFRS-I,BFRS-II} continues to provide a powerful tool for the 
theoretical interpretation of the corresponding $B$-factory data. The recent 
experimental progress allows us now to use only data where the results of the BaBar 
and Belle collaborations are in full agreement with each other. Interestingly, the 
resulting SM picture is very stable, with almost unchanged central values since the 
original analysis of 2003, and significantly reduced errors. 

In our new analysis, we pointed out that the branching ratio and direct CP 
asymmetry of the $B^0_d\to\pi^-K^+$ decay allow us to clarify the still unsatisfactory 
situation of the measurements of the direct CP violation in $B^0_d\to\pi^+\pi^-$:
\begin{itemize}
\item We predict ${\cal A}_{\rm CP}^{\rm dir}(B_d\to\pi^+\pi^-)=-0.24\pm0.04$, which
favours the BaBar result.
\item We extract $\gamma=\left(70.0^{+3.8}_{-4.3}\right)^\circ$, 
in agreement with the SM fits of the unitarity triangle. 
\end{itemize}
Moreover, we find hadronic parameters characterizing the $B\to\pi\pi$ system 
that show large CP-conserving strong phases, thereby establishing large
deviations from the naive factorization hypothesis.

The current status of the $B\to\pi K$ system can be summarized as follows:
\begin{itemize}
\item All modes with colour-suppressed EW penguins are found in excellent 
agreement with the SM.

\item The data for the $R_{\rm n,c}$ have moved quite a bit towards the SM predictions,
which are almost unchanged, thereby strongly reducing the ``$B\to\pi K$ puzzle" for the
CP-averaged branching ratios. 

\item On the other hand, the mixing-induced CP violation in $B^0_d\to\pi^0K_{\rm S}$ 
still looks puzzling, and can straightforwardly be accommodated through a modified 
EW penguin sector with a large, positive value of the CP-violating NP phase $\phi$.

\item The non-zero experimental value of $\Delta A$ in (\ref{Delta-A}) seems
to be caused by hadronic and not by NP effects. 
\end{itemize}
Unfortunately, we still cannot draw definite conclusions about the presence of
NP in the $B\to\pi K$ system (and other $b\to s$ penguin decays, such as
$B^0_d\to\phi K_{\rm S}$). It will be interesting to keep track of the picture
of these decays once the data improve further.

\vspace*{0.5truecm}

\noindent
{\it Acknowledgments}

\vspace*{0.2truecm}

\noindent
We would like to thank Andrzej Buras for a most enjoyable collaboration.
F.S. acknowledges financial support from the Deutsche Forschungsgemeinschaft (DFG) and from the 
``Bundesministerium f\"ur Bildung und Forschung (BMBF)" under contract 05HT6WOA.

\end{document}